\documentclass{IEEEtran}
\usepackage{cite}
\usepackage{amsmath,amssymb,amsfonts}
\usepackage{algorithmic}
\usepackage{graphicx}
\usepackage{textcomp}
\usepackage{tikz} 

\def\BibTeX{{\rm B\kern-.05em{\sc i\kern-.025em b}\kern-.08em
    T\kern-.1667em\lower.7ex\hbox{E}\kern-.125emX}}
\begin{document}
\title{Comments on `Design and Implementation of Model-Predictive Control With Friction Compensation on an Omnidirectional Mobile Robot'}
\author{Mohammad Biglarbegian, \IEEEmembership{Senior Member, IEEE}
\thanks{The author is with the School of Engineering, University of Guelph, ON, Canada (e-mail: mbiglarb@uoguelph.ca). }
}

\maketitle

\begin{abstract}
There are errors in the dynamics model in \cite{b1}. In addition, some details of the derivations and assumptions are missing in the paper. This letter was submitted to the IEEE Transactions on Mechatronics and although its merit was acknowledged, but was not finally approved to be published. I still think this work is worth disseminating and it is potentially very useful for students or practitioners. In this letter, (i) the assumptions made are presented and the governing dynamics with details are derived, and (ii) the correct equations followed by the correct component of the state-space model ($\boldsymbol{A}$) are given. 
\end{abstract}

\begin{IEEEkeywords}
Omni-directional mobile robots, governing dynamics, state-space model.
\end{IEEEkeywords}

\section{Introduction}
\label{sec:introduction}
There are errors in the stated dynamics of the omni-directional mobile robot in \cite{b1}. Also, some details of derivations were not given in the paper. The objective of this letter is to point out the errors, present the correct governing dynamics, and provide the details of the derivations of system dynamics. All the details and assumptions are provided followed by the correct governing equations and the correct $\boldsymbol{A}$ of the state-space model. These are very helpful for potential users interested in developing/using similar robots; particularly for developing controllers. 

This will be very beneficial to the future readers who (i) wish to adopt the dynamics of omni-directional wheeled robots, and (ii) are interested in the details of derivations and assumptions made to arrive at the dynamics and state-space equations. The parameters used in the system are first introduced in Table \ref{table-prm}.

\begin{table}[h!]
\caption{system's parameters}
\centering
\begin{tabular}{|c|c|}
\hline
 $\omega$ & robot angular speed  \\ \hline
 $T_{Mi}$ & $i$th motor's torque  \\ \hline
 $\omega_{Mi}$ & $i$th motor angular speed  \\ \hline
 $\omega_{Wi}$ & $i$th wheel angular speed  \\ \hline
 $v_{i}$ & $i$th motor translational velocity  \\ \hline
 $T_{Wi}$ & $i$th wheel's torque  \\ \hline
 $r_{i}$ & $i$th wheel's radius  \\ \hline
 $d$ & distance from the center of robot to each wheel  \\ \hline
 $N_{Mi}$ & number of teeth of gear of the $i$th motor  \\ \hline
 $N_{Wi}$ & number of teeth of the gear of the $i$th wheel  \\ \hline
 $l=\frac{N_{Wi}}{N_{Mi}}$ & gear ratio (reduction ratio)  \\ \hline
 $i_i$ & $i$th motor current  \\ \hline
 $u_{i}$ & $i$th input voltage  \\ \hline
 $R_{ai}$ & $i$th armature resistance (of the motor circuit)  \\ \hline
 $L_{ai}$ & $i$th armature inductance (of the motor circuit)  \\ \hline
 $K_{ti}, K_{vi}$ & $i$th motor constants  \\ \hline
\end{tabular}
\label{table-prm}
\end{table}

The traction force and torque on the $i$th wheel are respectively given by
\begin{align}
  f_i & = \frac{T_{Wi}}{r_{i}} \label{traction-force-1} \\
  T_{Wi} & = T_{Mi}.l \label{wheel-torque}
\end{align}
From \eqref{traction-force-1} and \eqref{wheel-torque}, we can express $f_i$ as
\begin{equation}
 f_i = \frac{T_{Mi}.l}{r_{i}} \label{traction-force-2}
\end{equation}
We also know the motor torque is proportional to the current, i.e.,
\begin{equation}
 T_{Mi} = K_{ti}.i_i  \label{torque-current}
\end{equation}
Using \eqref{traction-force-2} and \eqref{torque-current}, $f_i$ is given by
\begin{equation}
 f_i = \frac{K_{ti}.i_i.l}{r_i}  \label{traction-force-3}
\end{equation}

The armature circuit of each DC motor is modeled as
\begin{equation}
 u_{i} = R_{ai}.i_i + \frac{dL_{ai}}{dt}.i_i + K_{vi}.\omega_{Mi}  \label{motor-input-voltage-1}
\end{equation}

Paper \cite{b1} assumes the inductance of the armature circuit is small and hence neglected. Therefore,
\begin{equation}
 u_{i} \simeq R_{ai}.i_i + K_{vi}.\omega_{Mi}  \label{motor-input-voltage}
\end{equation}
From \eqref{motor-input-voltage}, the current can be expressed as
\begin{equation}
i_i= \frac{u_i- K_{vi}.\omega_{Mi}}{R_{ai}}   \label{current-1}
\end{equation}

The angular speeds of the wheel and motor are related to each other as follows:
\begin{equation}
\omega_{Mi} = \omega_{Wi}.{l}   \label{wheel-motor-speed-rlt}
\end{equation}
Assuming also there is small friction between the wheel and the ground, the relationship between the translational and rotational speed of each wheel is governed by
\begin{equation}
v_{i} = \omega_{Wi}.{r_i}   \label{gear-ratio}
\end{equation}
Using \eqref{wheel-motor-speed-rlt} and \eqref{gear-ratio} we can write \eqref{current-1} as
\begin{equation}
i_i = \frac{u_i-K_{vi}.\omega_{Wi}.l}{R_{ai}}= \frac{u_i-\frac{K_{vi}lv_i}{r_i}}{R_{ai}}  \label{current}
\end{equation}
The traction force in \eqref{traction-force-3} using \eqref{current} can be re-expressed as
\begin{equation}
f_i = \frac{K_{ti}l}{r_iR_{ai}} \big(u_i-\frac{K_{vi}lv_i}{r_i}\big)  \label{traction-force-4}
\end{equation}
Also it is assumed $K_{vi}=K_{ti}$, then
\begin{equation}
f_i = \frac{K_{ti}l}{r_iR_{ai}}u_i-\frac{K_{ti}^2l^2}{r_i^2R_{ai}}v_i  \label{traction-force}
\end{equation}
\begin{figure}[!t]
\centerline{\includegraphics[width=\columnwidth]{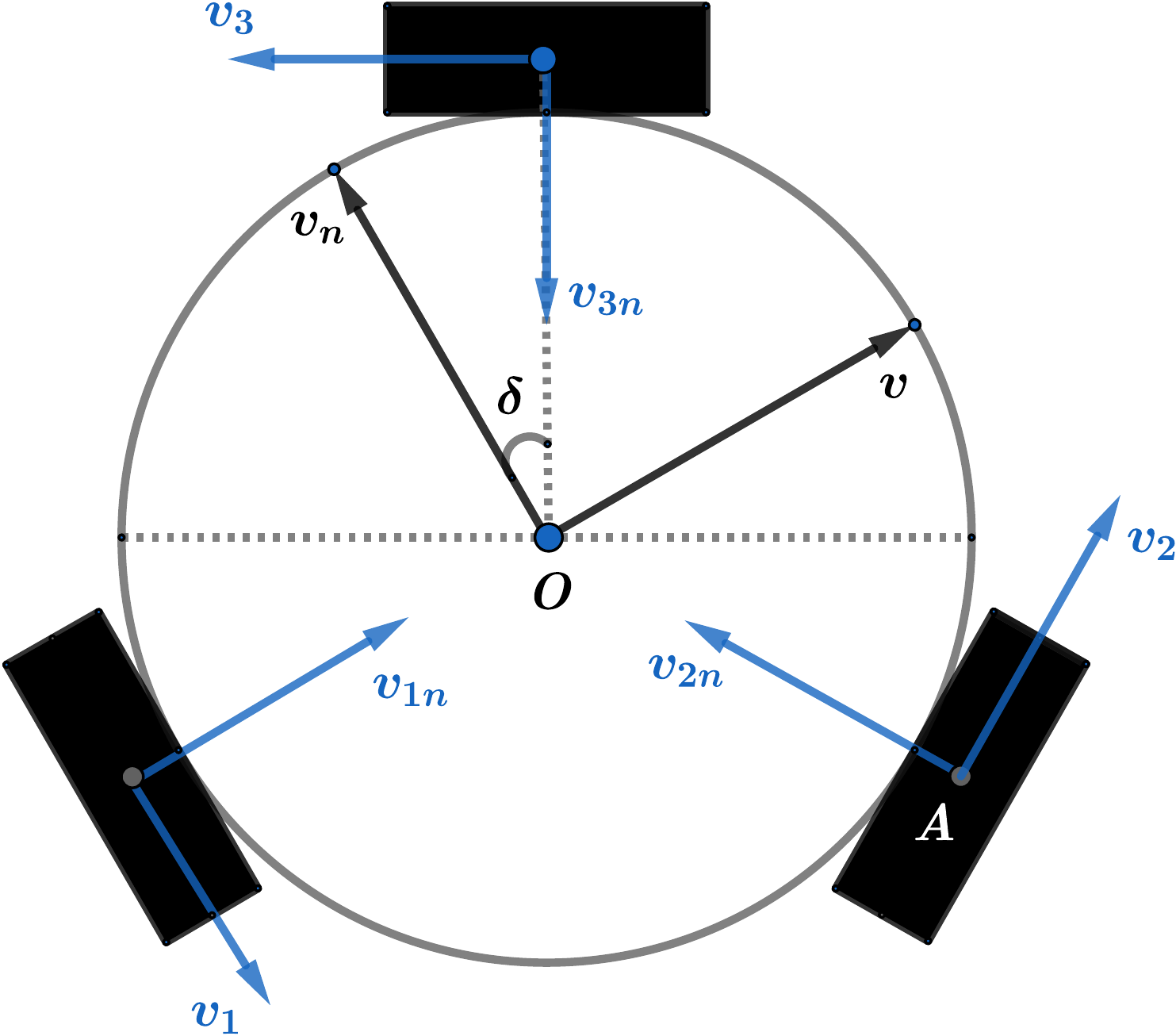}}
\caption{Robots and the coordinate systems used.}
\label{fig-1}
\end{figure}
\begin{figure}[!t]
\centerline{\includegraphics[width=\columnwidth]{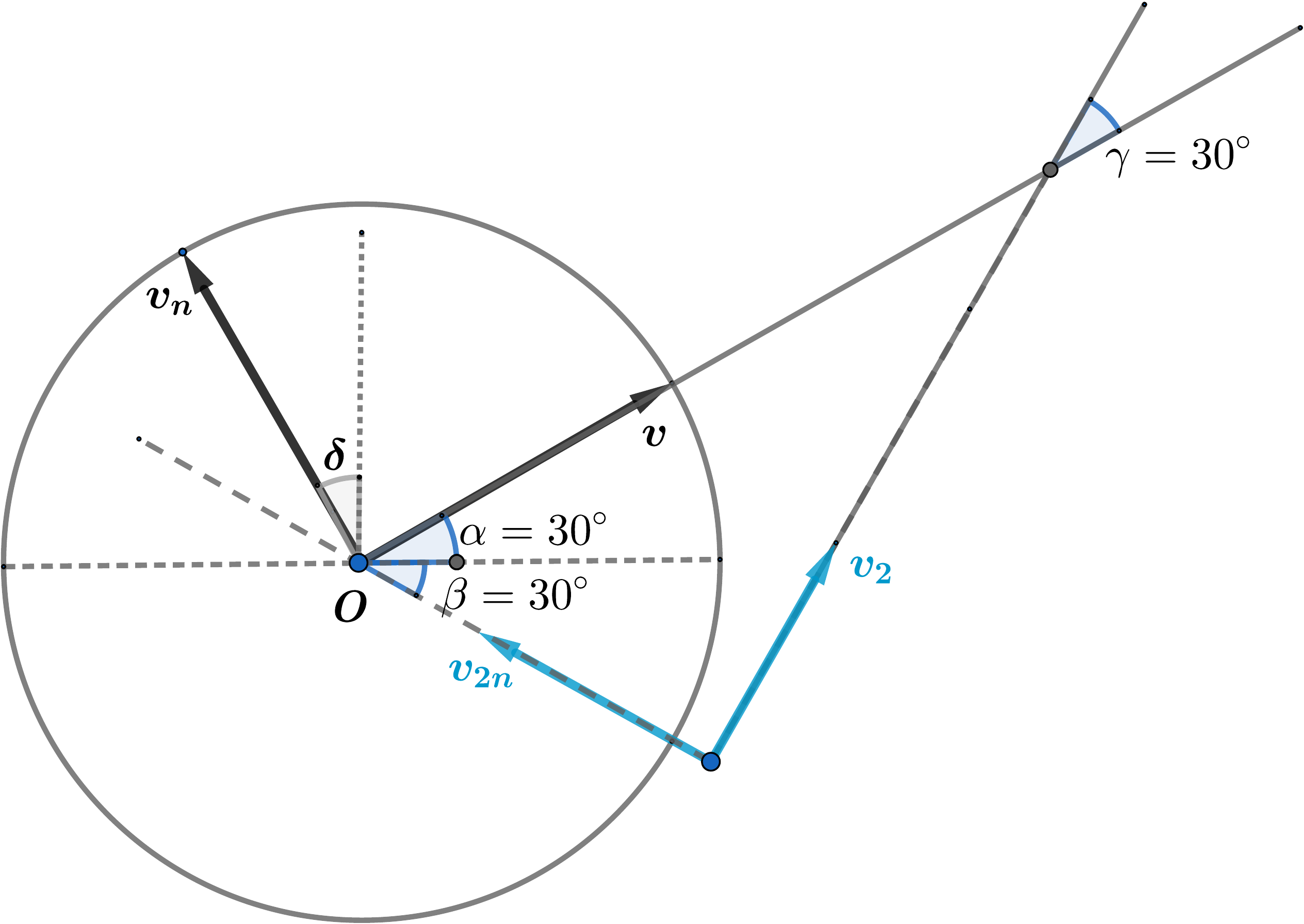}}
\caption{Geometrical relationship between coordinates.}
\label{fig-2}
\end{figure}

Note that the traction forces,  $f_i$ ($i=1,2,3$), depend on the velocities of the wheels. So, we first need to obtain $v_i$ and project each velocity along the axis of that wheel. Figure \ref{fig-1} shows the wheel coordinate systems ($v_i,v_{ni}$, for $i=1,2,3$) used including the robot body-fixed coordinate axes ($v$ and $v_{n}$). Note that in \cite{b1} it has been assumed that the angle between $v_{n}$ and the vertical axis is $\delta = 30^{\circ}$. In Figure \ref{fig-1} the translational speeds of each wheel are shown with respect to each wheel coordinate system. We focus our analysis on wheel $2$; for other wheels a very similar approach can be used. The translational velocity of wheel $2$ is given by
\begin{equation}
\vec{v}_{w2} = \vec{v}_0 + \overrightarrow{\omega} \times \overrightarrow{OA} \label{v2-1}
\end{equation}
The expressions for $\vec{v_0}$ and $\overrightarrow{\omega} \times \overrightarrow{OA}$ are given respectively as
\begin{equation}
\vec{v_0} = v \hat{v} + v_n \hat{v_n}
\end{equation}
where $\hat{v}$ and $\hat{v_n}$ are unit vectors of the robot coordinate system (body-fixed), and
\begin{equation}
\overrightarrow{\omega} \times \overrightarrow{OA} = \omega d \hat{v}_{2}
\end{equation}
where $\hat{v}_{2}$ is the unit vector along the direction of the wheel. Knowing that
\begin{align}
\hat{v} = cos\delta\hat{v}_2 - sin\delta\hat{v}_{2n} \\
\hat{v}_{n} = sin\delta\hat{v}_2 + cos\delta\hat{v}_{2n}
\end{align}
where $\hat{v}_{2n}$ is the unit vector perpendicular to $\hat{v}_2$, $v_2$ in \eqref{v2-1} can be expressed in terms of $\hat{v}_{2}$ and $\hat{v}_{2n}$ as follows:
\begin{equation} \label{abc}
\begin{split}
\vec{v}_{w2} & = vcos\delta\hat{v}_2 - vsin\delta\hat{v}_{2n}\\
& \quad + v_nsin\delta\hat{v}_2 + v_ncos\delta\hat{v}_{2n} +  \omega d \hat{v}_2 \\
\end{split}
\end{equation}
The velocity of the wheel along the $\hat{v}_{2}$ axis is $vcos\delta+v_nsin\delta+\omega d$. On the other hand, we know this velocity should be $v_2$, i.e.,
\begin{equation}
v_2= vcos\delta+v_nsin\delta+\omega d \label{v2}
\end{equation}

Similarly, for the first and third wheels we have
\small{
\begin{align}
v_1 & = -v_n,                                   \label{v1} \\
v_3 & = -vcos\delta+v_nsin\delta+\omega d       \label{v3}
\end{align}
}
Therefore,
\begin{equation}
\begin{bmatrix}
v_1 \\
v_2 \\
v_3
\end{bmatrix} =
\begin{bmatrix}
0 & -1 & 0 \\
cos\delta & sin\delta & d  \\
-cos\delta & sin\delta & d
\end{bmatrix}
\begin{bmatrix}
v \\
v_n \\
\omega
\end{bmatrix} \label{v1,v2,v3}
\end{equation}

Using \eqref{v1,v2,v3} and assuming $K_{ti}=K_{vi}=K_{t}$, we can calculate the traction forces as follows:
\begin{align}
f_1 & = \frac{K_{t}l}{rR_{a}}u_1+ \frac{K_{t}^2l^2}{r^2R_{a}}v_n \label{f_1} \\
f_2 & = \frac{K_{t}l}{rR_{a}}u_2- \frac{K_{t}^2l^2}{r^2R_{a}} (vcos\delta+v_nsin\delta+\omega d) \label{f_2} \\
f_3 & = \frac{K_{t}l}{rR_{a}}u_3- \frac{K_{t}^2l^2}{r^2R_{a}} (-vcos\delta+v_nsin\delta+\omega d) \label{f_3}
\end{align}

The tangential and normal forces as well as the torque acting on the robot along the $v$ and $v_n$ axes are given by
\begin{align}
F_v & = cos\delta(f_2-f_3) \label{F-v} \\
F_{vn} & = -f_1+ sin\delta(f_2+f_3) \label{F_n-v} \\
\Gamma & = b(f_1+f_2+f_3) \label{Torque}
\end{align}
It is worth noting that $b$ is the distance from the robot center to the center of each wheel and equal to $d$. In this paper, $b$ was also used to be consistent with the notion used in \cite{b1}. 

Using \eqref{f_1}, \eqref{f_2}, and \eqref{f_3} results in
\begin{align}
F_v  & = cos\delta \left[ \frac{K_{t}l}{rR_{a}}\Big(u_2-u_3 \Big)- \frac{K_{t}^2l^2}{r^2R_{a}}\left(v\sqrt3\right)\right] \label{F-v: p2}  \\
\begin{split}
F_{vn} & = \frac{K_{t}l}{rR_{a}}\left(-u_1+sin\delta(u_2+u_3)\right)\\
& \quad - \frac{K_{t}^2l^2}{r^2R_{a}}v_n - \frac{K_{t}^2l^2}{r^2R_{a}} sin\delta(2v_nsin\delta+2\omega d)  \\
\end{split}\label{F-vn: p2}\\
\Gamma & = b\left[\frac{K_{t}l}{rR_{a}}\left(u_1+u_2+u_3\right)+ \frac{K_{t}^2l^2}{r^2R_{a}}v_n - \frac{K_{t}^2l^2}{r^2R_{a}} \big(2v_nsin\delta+2\omega d\big)\right] \label{Torque: p2}
\end{align}

Eqs. \eqref{F-v: p2}-\eqref{Torque: p2} can be further simplified to
\begin{align}
F_v  & =  cos\delta \frac{K_{t}l}{rR_{a}}\Big(u_2-u_3 \Big)- \frac{3}{2}\frac{K_{t}^2l^2}{r^2R_{a}}v  \label{correct F-v} \\
\begin{split}
F_{vn} & = \frac{K_{t}l}{rR_{a}}\left(-u_1+sin\delta(u_2+u_3)\right)- \frac{K_{t}^2l^2}{r^2R_{a}}v_n \\
& \quad - \frac{K_{t}^2l^2}{r^2R_{a}}\left(\frac{v_n}{2}+\omega d\right)  \\
&  = \frac{K_{t}l}{rR_{a}}\left(-u_1+sin\delta(u_2+u_3)\right)-\frac{K_{t}^2l^2}{r^2R_{a}}\left(\frac{3v_n}{2}+\omega d\right) \label{correct F_n-v} \\
\end{split}\\
\begin{split}
\Gamma & = b\left[\frac{K_{t}l}{rR_{a}}\left(u_1+u_2+u_3\right)+ \frac{K_{t}^2l^2}{r^2R_{a}}v_n - \frac{K_{t}^2l^2}{r^2R_{a}} \left(v_n+2\omega d\right)\right] \\
&  = b\left[\frac{K_{t}l}{rR_{a}}\left(u_1+u_2+u_3\right)- \frac{K_{t}^2l^2}{r^2R_{a}} \left(2\omega d\right)\right] \label{correct Torque}
\end{split}
\end{align}

One can also substitute $\delta = 30^{\circ}$ yielding
\begin{align}
F_v  & =  \frac{\sqrt 3}{2} \frac{K_{t}l}{rR_{a}}\Big(u_2-u_3 \Big)- \frac{3}{2}\frac{K_{t}^2l^2}{r^2R_{a}}v  \label{correct_with delta30 F-v} \\
\begin{split}
F_{vn} & = \frac{K_{t}l}{rR_{a}}\left(-u_1+\frac{1}{2}(u_2+u_3)\right)-\frac{K_{t}^2l^2}{r^2R_{a}}\left(\frac{3v_n}{2}+\omega d\right) \label{correct_with delta30 F_n-v} \\
\end{split}\\
\begin{split}
\Gamma  = b\left[\frac{K_{t}l}{rR_{a}}\left(u_1+u_2+u_3\right)- \frac{K_{t}^2l^2}{r^2R_{a}} \left(2\omega d\right)\right] \label{correct_with delta30 Torque}
\end{split}
\end{align}

Eqs. \eqref{correct F-v}-\eqref{correct Torque} or \eqref{correct_with delta30 F-v}-\eqref{correct_with delta30 Torque} are the correct expressions for forces and torque on the robot.\newline \newline
\textbf{Note 1:} in \eqref{correct F_n-v} the coefficient of $\omega$ is non-zero and it is $-\frac{K_{t}^2l^2}{r^2R_{a}}d$. Paper \cite{b1} has missed that coefficient.\\ \newline\textbf{Note 2:} in \eqref{correct Torque} the coefficient of $\omega$ is $-2\frac{K_{t}^2l^2}{r^2R_{a}}b^2$ and once divided by $I_n$ (to give a portion of $\frac{d\omega(t)}{dt}$ required to calculate the state-space) is $-2\frac{K_{t}^2l^2}{r^2R_{a}.I_n}b^2$ which is different when compared to the corresponding component of $\boldsymbol{A}$ given in \cite{b1}. The errors in paper \cite{b1} are the components of both $F_{vn}$ and $\Gamma$ in $\boldsymbol{A}$.

Therefore, the correct matrix $\boldsymbol{A}$ of the state-space model is:
\begin{equation}
\boldsymbol{A}=
\begin{bmatrix}
\frac{-3}{2M}\frac{K_{t}^2l^2}{r^2R_{a}}-\frac{B_v}{M} & 0 & 0 \\
0 & -\frac{3}{2M}\frac{K_{t}^2l^2}{r^2R_{a}}-\frac{B_{vn}}{M}  & -\frac{K_{t}^2l^2}{r^2R_{a}.M}d  \\
0 & 0 & -2\frac{K_{t}^2l^2b^2}{r^2R_{a}I_n}-\frac{B_{\omega}}{I_n}
\end{bmatrix} \label{A matrix}
\end{equation}

As stated, the term $-\frac{K_{t}^2l^2}{r^2R_{a}}d$ is missing in the expression of $F_{vn}$ in paper [1]. Depending on the values of $K_t,l,r,R_a,d$  this term is not negligible. In fact, when $r$ or $R_a$ is small/very small, this term can become large/very large. This term plays an important role especially when $\omega$ is non-zero and will affect the controller's performance especially in the lateral direction. 

Moreover, the correct expression for computing the torque $-2\frac{K_{t}^2l^2}{r^2R_{a}.I_n}b^2$ which has a noticeable different magnitude than in \cite{b1}. This will affect the equation for computing the torque in \eqref{torque-current} and consequently the angular acceleration, which in effect changes both $\omega$ and then ultimately $F_{vn}$. 

As can be seen from \eqref{correct F-v}-\eqref{correct Torque}, the equations of motion are coupled with each other and hence a missing or/and incorrect term will result in incorrect dynamics of the robot; specially the two components of lateral and angular. Particularly when model-based controllers such as MPC are utilized, since model-based controllers rely heavily on the robot model, incorrect dynamics will in turn have a negative consequence on the controller's performance (e.g. being poor or an unstable system).


\begin{thebibliography}{00}

\bibitem{b1} J. C. L. Barreto S. and A. G. S. Concei\c{c}\~{a}o and C. E. T. D\'{o}rea and L. Martinez and E. R. de Pieri, ``Design and Implementation of Model-Predictive Control With Friction Compensation on an Omnidirectional Mobile Robot,'' in \emph{IEEE/ASME Trans. on Mechatronics,} vol. 19, no. 2, pp. 467--476, Apr. 2014.

\end{thebibliography}
\end{document}